\documentstyle[11pt]{article}
\author{G. Grignani and P. Sodano\\
{\em Department of Physics and INFN, University of Perugia, Perugia, Italy}
\\ \\ C. A. Scrucca\\
{\em International School for Advanced Studies and INFN, Trieste, Italy}
\\  \\}

\title{QUANTUM STATES OF TOPOLOGICALLY MASSIVE ELECTRODYNAMICS AND GRAVITY
}

\catcode`\@=11
\newdimen\ex@
\def\rightarrowfill@#1{$#1\m@th\mathord-\mkern-6mu\cleaders
 \hbox{$#1\mkern-2mu\mathord-\mkern-2mu$}\hfill
 \mkern-6mu\mathord\rightarrow$}
\def\leftarrowfill@#1{$#1\m@th\mathord\leftarrow\mkern-6mu\cleaders
 \hbox{$#1\mkern-2mu\mathord-\mkern-2mu$}\hfill\mkern-6mu\mathord-$}
\def\leftrightarrowfill@#1{$#1\m@th\mathord\leftarrow\mkern-6mu\cleaders
 \hbox{$#1\mkern-2mu\mathord-\mkern-2mu$}\hfill
 \mkern-6mu\mathord\rightarrow$}
\def\overrightarrow{\mathpalette\overrightarrow@}
\def\overrightarrow@#1#2{\vbox{\ialign{##\crcr\rightarrowfill@#1\crcr
 \noalign{\kern-\ex@\nointerlineskip}$\m@th\hfil#1#2\hfil$\crcr}}}

\def\overleftarrow{\mathpalette\overleftarrow@}
\def\overleftarrow@#1#2{\vbox{\ialign{##\crcr\leftarrowfill@#1\crcr
 \noalign{\kern-\ex@\nointerlineskip}$\m@th\hfil#1#2\hfil$\crcr}}}
\def\overleftrightarrow{\mathpalette\overleftrightarrow@}
\def\overleftrightarrow@#1#2{\vbox{\ialign{##\crcr\leftrightarrowfill@#1\crcr
 \noalign{\kern-\ex@\nointerlineskip}$\m@th\hfil#1#2\hfil$\crcr}}}
\catcode`\@=\active
\def\dint{\displaystyle \int }
\def\dfrac#1#2{{\displaystyle {#1 \over #2}}}

\advance\textheight by \topskip
\begin{document}

\maketitle
\hspace{10.cm} DFUPG-120/96\\

\hspace{10.cm} hep-th 9602127\\

\begin{abstract}
The free quantum states of topologically massive electrodynamics and gravity 
in 2+1 dimensions, are explicitly found. It is shown that in both theories the
states are described by infrared-regular polarization tensors containing a
regularization phase which depends on the spin. This is done by explicitly 
realizing the quantum algebra on a functional Hilbert space and by finding
the Wightman function to define the scalar product on such a Hilbert space.
The physical properties of the states are analyzed defining creation and
annihilation operators.

For both theories, a canonical and covariant quantization procedure is 
developed. The higher order derivatives in the gravitational lagrangian are 
treated by means of a preliminary Dirac procedure.

The closure of the Poincar\'e algebra is guaranteed by the infrared-finiteness
of the states which is related to the spin of the excitations through the
regularization phase. Such a phase may have interesting physical consequences.

\end{abstract}

\section{Introduction}

\quad \ \thinspace 
The topological mass arising from the Chern-Simons term in 2+1-dimensional
topologically massive gauge theories \cite{sie,sch,djt}, provides, at the 
quantum level, an infrared cut-off which seems to cure the infrared 
problem without disturbing 
the ultraviolet or gauge properties of these theories. However, as it was 
recognized already 
in the seminal work by Deser, Jackiw and Templeton \cite{djt},
a few delicate points, concerning 
the infrared behavior of these theories, need a careful treatment. In 
particular the closure of the Poincar\'e algebra, seems to be 
subordinate to a particular definition of the phase of
the field operators \cite{djt}.

In this paper we shall perform a careful analytical derivation of the free 
quantum states of topologically massive planar theories devoting 
particular attention to the 
treatment of the infrared ambiguities.
The exact knowledge of the states of these theories leads
to that of the polarization tensors. 
For electrodynamics, the states are known but
infrared-ambiguous \cite{2}, whereas for gravity, they have yet not been
studied. 
We shall show that in both theories the states are described by infrared 
regular polarization tensors containing a regularization phase which 
depends on the spin. Such a phase was first introduced in Ref.\cite{djt}. 

For electrodynamics, an infrared modification of the polarization vector
of physical photons leads to interesting consequences in phenomenological
applications of the interacting theory with fermionic matter. In
particular, the correct form of the polarization vector might help in 
proving the  conjectured existence of fermion-photon bound-states\cite{3}.
In the gravitational case, the calculation of the polarization tensor of 
physical gravitons
might be considered as a first step for the understanding of some 
questions arising, at the quantum level, in the relation between the first 
and second order formulations of the theory\cite{4}.

In the following we shall use a general procedure \cite{5,ros} to construct 
the Hilbert space of the physical states and its scalar 
product. The latter is defined in terms of the pertinent 
two-points Wightman function 
which we shall explicitly compute for both topologically massive 
electrodynamics and gravity.
Such a procedure has been succesfully used also in constructing Fock 
spaces with generalized statistics\cite{min}.

We shall perform a canonical analysis keeping into account both the 
constraints related to gauge invariance and those, arising in the 
gravitational case, due to the higher order derivatives of the lagrangian.
In order to work always with well defined quantities one needs a careful 
definition of the action of the operators on the quantum states; 
this is achieved by introducing a suitable set of test functions. 

We analyse topologically massive electrodynamics and gravity  
simultaneously, taking as lagrangians \begin{equation}
{\cal L}_{ELEC}{\cal =}-\frac 14F_{\mu \nu }F^{\mu \nu }+\frac \mu  
2{\cal E}^{\mu \nu \rho }\partial _\mu A_\nu A_\rho \;\,, 
\end{equation}
\begin{equation}
{\cal L}_{GRAV}=\sqrt{g}R+\frac 1{2\mu }{\cal E}^{\mu \lambda \nu }\Gamma
^\rho {}_{\lambda \sigma }\left( \partial _\mu \Gamma ^\sigma {}_{\nu \rho
}+\frac 23\Gamma ^\sigma {}_{\mu \xi }\Gamma ^\xi {}_{\nu \rho }\right)
\;\,. 
\end{equation}
 
In Section 2 we shall define the canonical variables and brackets in 
electrodynamics and linearized gravity, using in both cases a covariant 
gauge choice, Lorentz and Landau respectively. For linearized 
gravity a preliminary Dirac procedure is needed. This allows us to find the 
canonical brackets that are consistent with the second class constraints 
introduced when enlarging the configuration space to the time derivatives of 
some components of the metric \cite{whit}.

In Section 3 we shall derive the two-points Wightman functions by solving 
the Cauchy problem arising from the equations of motion and the equal-time
commutators. For linearized gravity the derivation turns out to be quite 
difficult since it involves a third order Cauchy problem.  Nevertheless, 
for both theories, the explicit solution of the Cauchy problem is provided.

In Section 4 we construct the 
Hilbert space of the physical states deriving  the 
expression for the one-particle states. The polarization tensors 
constructed in this section are well defined in the infrared. As a 
consequence, the states we exhibit are the only possible states allowing 
for the weak closure of the 
Poincar\'e algebra.

In Section 5 we analyse the physical properties of the quantum states, 
their mass and spin. To achieve this result we perform a normal mode 
expansion obtained in terms of the Wightman function.

Section 6 is devoted to some concluding remarks.

\section{Canonical brackets}

\quad \ \thinspace 
Let us start by introducing the canonical variables and by finding the 
canonical brackets of
topologically massive electrodynamics and linearized gravity. We 
choose, in the framework of an indefinite metric Fock space, covariant
gauges to maintain explicit Lorentz invariance.

\subsection{Electrodynamics}

\quad \ \thinspace
The gauge-fixed lagrangian is, in the Lorentz gauge $\partial _\mu A^{\mu}=0$, 
\begin{equation}
{\cal L=}-\frac 14F_{\mu \nu }F^{\mu \nu }+\frac \mu 2{\cal E}^{\mu \nu \rho
}\partial _\mu A_\nu A_\rho -\frac 1{2\xi }\left( \partial _\mu A^\mu
\right) ^2\;\,,
\label{qed} 
\end{equation}
leading to the equations of motion
\begin{equation}
\label{MotoE}\Box A^\mu +\left( \xi -1\right) \partial ^\mu \partial _\nu
A^\nu +\mu {\cal E}^{\mu \nu \rho }\partial _\nu A_\rho =0\;\,. 
\end{equation}

The lagrangian (\ref{qed}) is regular, i.e. $\det \left(
\partial ^2{\cal L} / \partial \left( \partial ^0A^\mu \right) \partial
\left( \partial ^0A^\nu \right) \right) \neq 0$, and the canonical momentum 
\begin{equation}
\pi ^\mu =\frac{\partial {\cal L}}{\partial \left( \partial _0A_\mu  
\right) }
=F^{\mu 0}+\frac \mu 2{\cal E}^{0\mu \nu }A_\nu -\xi \eta ^{\mu 0}\partial
_\nu A^\nu 
\end{equation}
can be uniquely inverted to obtain the velocities.
The hamiltonian version of the theory
is constructed in the usual way; the only non-vanishing canonical
Poisson bracket is 
\begin{equation}
\left\{ A^\mu \left( \vec x,t\right) ,\pi ^\nu \left( \vec y,t\right)
\right\} =\eta ^{\mu \nu }\delta ^2\left( \vec x-\vec y\right) \;\,.
\label{qedbra}
\end{equation}
The energy-momentum and angular-momentum tensors have the
usual form 
\begin{equation}
T^{\mu \nu }=\pi ^{\rho (\mu )}\partial ^\nu A_\rho -\eta ^{\mu \nu } 
{\cal L} \;\,, 
\end{equation}
\begin{equation}
M^{\alpha \mu \nu }=x^\mu T^{\alpha \nu }-x^\nu T^{\alpha \mu }+\pi ^{\rho
(\alpha )}\left( \Sigma ^{\mu \nu }{}\right) _{\rho \tau }A^\tau \;\,, 
\end{equation}
where 
\begin{equation}
\pi ^{\rho (\mu )}=\frac{\partial {\cal L}}{\partial \left( \partial _\mu
A_\rho \right) }\;\,, 
\end{equation}
and 
\begin{equation}
\left( \Sigma ^{\mu \nu }{}\right) _{\rho \tau }=\delta _\rho ^\mu \delta
_\tau ^\nu -\delta _\tau ^\mu \delta _\rho ^\nu 
\end{equation}
is the spin matrix of the field $A^\mu $.

The quantum version of the theory is obtained using the correspondence
principle 
\begin{equation}
\left\{ A,B\right\} \rightarrow -i\left[ A,B\right] 
\end{equation}
and defining the physical states as the ones destroyed by the annihilation 
part of $\partial\cdot A$
\begin{equation}
\left( \partial _\mu A^\mu \right) ^{-}\left| \Phi _f\right\rangle =0\;\,. 
\end{equation}

\subsection{Gravity}

\quad \ \thinspace 
To perform a canonical analysis of the topologically massive gravity
we consider its linearized version, 
$g^{\mu \nu }=\eta^{\mu \nu }+kh^{\mu\nu}$, 
with the Minkowski signature $\hbox{diag}\left(\eta_{\mu\nu}\right) 
=\left(+,-,-\right)$.
The linearized lagrangian is invariant under the abelian gauge transformations
\begin{equation}
\delta h_{\mu\nu} = \partial_\mu\xi_\nu+\partial_\nu\xi_\mu\ \ .
\label{gau}
\end{equation}
To fix the gauge we then select the Landau gauge $\partial
_\mu h^{\mu \nu }=0$. The linearized lagrangian is \cite{djt}
\begin{equation}
\begin{array}{l}
{\cal L}=-\dfrac 14\left[ \left( \partial _\rho h_{\mu \nu }\right) \left(
\partial ^\rho h^{\mu \nu }\right) -\left( \partial _\mu h\right) \left(
\partial ^\mu h\right) +2\left( \partial ^\nu h\right) \left( \partial ^\mu
h_{\mu \nu }\right) +2\left( \partial ^\mu h_{\mu \nu }\right) \left(
\partial _\rho h^{\rho \nu }\right) \right] \medskip\  \\ \qquad +\dfrac
1{4\mu }{\cal E}_{\mu \alpha \beta }\left( \partial ^\alpha \partial _\nu
h_\rho {}^\beta -\partial ^\alpha \partial _\rho h_\nu {}^\beta \right)
\partial ^\nu h^{\mu \rho }-\dfrac 1{2\xi }\left( \partial _\mu h^{\mu
\lambda }\right) \left( \partial _\nu h^\nu {}_\lambda \right) \;\,. 
\label{gra}
\end{array}
\end{equation}

Since (\ref{gra}) contains second order derivatives, the equations of
motion are third order 
\begin{equation}
\label{MotoG} 
\begin{array}{l}
\left\{ \Box h^{\mu \nu }+\partial ^\mu \partial ^\nu h+\left( \xi -1\right)
\left( \partial ^\nu \partial _\alpha h^{\alpha \mu }+\partial ^\mu \partial
_\alpha h^{\alpha \nu }\right) -\eta ^{\mu \nu }\left( \Box h-\partial
_\alpha \partial _\beta h^{\alpha \beta }\right) \right. 
\medskip\  \\ \left. -\dfrac 1{2\mu }\left[ {\cal E}^{\mu \alpha \beta
}\partial _\alpha \left( \Box h_\beta {}^\nu -\partial ^\nu \partial
_\lambda h^\lambda {}_\beta \right) +{\cal E}^{\nu \alpha \beta }\partial
_\alpha \left( \Box h_\beta {}^\mu -\partial ^\mu \partial _\lambda
h^\lambda {}_\beta \right) \right] \right\} =0\;\,. 
\end{array}
\end{equation}
In general the order of the equations
of motion should be twice that of the lagrangian; the fact that this does not
occur here is a first evidence of the constrained nature of this theory.
In fact, the lagrangian is singular, i.e.
 $\det \left({\partial ^2{\cal L}}/{\partial \left( \partial ^0\partial
^0A^\mu \right) \partial \left( \partial ^0\partial ^0A^\nu \right)}\right)
=0$.
In Eq. (\ref{gra}) all the dependence from the second order
derivatives lies in the Chern-Simons lagrangian, whose only  
non-vanishing terms are \begin{equation}
{\cal L}_{CS}=\frac 1{4\mu }{\cal E}_{ij}\left( \partial ^kh^{i0}-\partial
^0h^{ik}\right) \left( \partial ^0\right) ^2h_k{}^j\;\,. 
\label{lcs}
\end{equation}
Therefore, only the space-space components of $h^{\mu \nu }$ 
have higher order dynamics.

To extend the hamiltonian formulation to this case, one needs to 
decouple the second order time-derivatives,  so that the dynamics
becomes first order in time. This is done by defining an additional
canonical variable for the time-derivative of each variable having
higher order dynamics. One can then extend the definition of the canonical 
momentum so that the formal structure of the canonical Legendre
transformation is maintained \cite{whit}. This procedure has been  
applied to topologically
massive gravity in Ref. \cite{ben, bar}, whereas the canonical  
formulation of the
full nonlinear theory has been analyzed in Ref. \cite{des}.  

At variance with Ref. \cite{bar},
it is clear from (\ref{lcs}) that it suffices to take as independent canonical 
variables $h^{\mu\nu }$ and $k^{ij}=\partial^0h^{ij}$,
together with their respective
conjugate momenta, defined as 
\begin{equation}
\pi ^{\mu \nu }=\frac{\partial {\cal L}}{\partial \left( \partial _0h_{\mu
\nu }\right) }-2\partial _\rho \frac{\partial {\cal L}}{\partial \left(
\partial _\rho \partial _0h_{\mu \nu }\right) }+\partial _0\frac{\partial 
{\cal L}}{\partial \left( \partial _0\partial _0h_{\mu \nu }\right) }\;\,, 
\end{equation}
\begin{equation}
s^{ij}=\frac{\partial {\cal L}}{\partial \left( \partial _0\partial
_0h_{ij}\right) }\;\,. 
\end{equation}
In terms of the above variables,
everything works essentially as usual,
except for the fact that the phase space has been extended. 
In particular, the only
non-vanishing canonical Poisson brackets are 
\begin{equation}
\left\{ h^{\mu \nu }\left( \vec x,t\right) ,\pi ^{\alpha \beta }\left( \vec
y,t\right) \right\} =\frac 12\left( \eta ^{\mu \alpha }\eta ^{\nu \beta
}+\eta ^{\mu \beta }\eta ^{\nu \alpha }\right) \delta ^2\left( \vec x-\vec
y\right) \;\,, 
\end{equation}
\begin{equation}
\left\{ k^{ij}\left( \vec x,t\right) ,s^{mn}\left( \vec y,t\right) \right\}
=\frac 12\left( \eta ^{im}\eta ^{jn}+\eta ^{in}\eta ^{jm}\right) \delta
^2\left( \vec x-\vec y\right) \;\,, 
\end{equation}
and the energy-momentum and angular-momentum tensors become 
\begin{equation}
T^{\mu \nu }=\pi ^{\rho \tau (\mu )}\partial ^\nu h_{\rho \tau }+s^{ij(\mu
)}\partial ^\nu k_{ij}-\eta ^{\mu \nu }{\cal L}\;\,, 
\end{equation}
\begin{equation}
M^{\mu \alpha \beta }=x^\alpha T^{\mu \beta }-x^\beta T^{\mu \alpha }+\pi
^{\rho \tau (\mu )}\left( \Sigma ^{\alpha \beta }\right) _{\rho \tau
}{}^{\lambda \sigma }h_{\lambda \sigma }+s^{ij(\mu )}\left( \Sigma ^{\alpha
\beta 0\lambda }\right) _{ij}{}^{\rho \tau }\partial _\lambda h_{\rho \tau
}\;\,, 
\end{equation}
where 
\begin{equation}
\pi ^{\mu \nu (\alpha )}=\frac{\partial {\cal L}}{\partial \left( \partial
_\alpha h_{\mu \nu }\right) }-2\partial_\rho \frac{\partial {\cal L}}{%
\partial \left( \partial _\rho \partial_\alpha h_{\mu \nu }\right) }%
+\partial _0\frac{\partial {\cal L}}{\partial \left( \partial _0\partial
_\alpha h_{\mu \nu }\right) }\;\,, 
\end{equation}
\begin{equation}
s^{ij(\alpha )}=\frac{\partial {\cal L}}{\partial \left( \partial _\alpha
\partial _0h_{ij}\right) }\;\,, 
\end{equation}
and 
\begin{equation}
\left( \Sigma ^{\mu \nu }\right) ^{\rho \tau \alpha \beta }=\frac 12\left[
\eta ^{\mu \rho }\eta ^{\nu \alpha }\eta ^{\tau \beta }+\eta ^{\mu \tau
}\eta ^{\nu \beta }\eta ^{\rho \alpha }+\left( \alpha \leftrightarrow \beta
\right) -\left( \mu \leftrightarrow \nu \right) \right] \;\,, 
\end{equation}
\begin{equation}
\left( \Sigma ^{\mu \nu 0\lambda }\right) ^{ij\alpha \beta }=\left( \Sigma
^{\mu \nu }\right) ^{ij\alpha \beta }\eta ^{\lambda 0}-\frac 12\left( \eta
^{\mu 0}\eta ^{\nu \lambda }-\eta ^{\nu 0}\eta ^{\mu \lambda }\right) \left(
\eta ^{i\alpha }\eta ^{j\beta }+\eta ^{i\beta }\eta ^{j\alpha }\right) 
\end{equation}
are the spin matrices of the fields $h^{\mu \nu }$ and $k^{ij}$.

A straightforward calculation gives for the canonical momenta
\begin{equation}
\pi ^{00}=-\xi \partial _0h^{00}+\left( \frac 12-\xi \right) \partial
_ih^{i0}\;\,, 
\end{equation}
\begin{equation}
\begin{array}{l}
\pi ^{0i}=-\dfrac 12\left[ \xi \partial _0h^{0i}+\left( \xi -1\right)
\partial _jh^{ji}+\frac 12\partial ^ih\right] -\dfrac 1{4\mu } 
{\cal E}_{mn}\left[ \partial ^m\partial ^ih^{0n}-\partial ^mk^{in}\right] 
\medskip\  \\ \qquad \;\;-\dfrac 1{8\mu }{\cal E}_m{}^i\left[ \partial
_jk^{mj}-\partial ^j\partial _jh^{m0}\right] \;\,, 
\end{array}
\end{equation}
\begin{equation}
\begin{array}{l}
\pi ^{ij}=-\dfrac 12\left[ k^{ij}-\eta ^{ij}k^l{}_l+\eta ^{ij}\partial
_kh^{k0}{}\right] 
\medskip\  \\ \qquad \;\;+\dfrac 1{8\mu }\left[ 
{\cal E}_k{}^j\left( 2\partial _0k^{ki}-2\partial ^i\partial
_0h^{k0}-2\partial ^k\partial _0h^{i0}\right. \right. \medskip\  \\ \qquad
\qquad \qquad \quad \;\left. \left. +2\partial ^k\partial ^ih^{00}+\partial
^l\partial _lh^{ki}-\partial ^i\partial _lh^{kl}\right) +\left(
i\leftrightarrow j\right) \right] \;\,, 
\label{pij}
\end{array}
\end{equation}
\begin{equation}
s^{ij}=-\frac 1{8\mu }\left[ {\cal E}_k{}^j\left( k^{ki}-\partial
^ih^{k0}\right) +\left( i\leftrightarrow j\right) \right] \;\,. 
\label{sij}
\end{equation}
From the trace of Eq. (\ref{pij}) and Eq. (\ref{sij}),
follow the constraints
\begin{equation}
\Lambda =\pi ^l{}_l-\frac 12k^l{}_l+\partial _kh^{k0}{}+\frac 1{4\mu }%
{\cal E}_{ki}{}\partial ^i\partial _lh^{kl}\approx 0\;\,, 
\label{lam}
\end{equation}
\begin{equation}
{\cal O}^{ij}=s^{ij}+\frac 1{8\mu }\left[ {\cal E}_k{}^j\left(
k^{ki}-\partial ^ih^{k0}\right) +\left( i\leftrightarrow j\right) \right]
\approx 0\;\,. 
\label{oij}
\end{equation}
This shows that the lagrangian remains singular even after 
the gauge-fixing. This is due to the fact that the constraints
(\ref{lam}), (\ref{oij}), are not related to the gauge invariance  
(\ref{gau}), but to the
enlargement of the phase space produced by the introduction of the new
variables $k_{ij}$. These constraints are in fact second class.
The canonical Poisson brackets are
incompatible with these constraints 
\footnote{The meaning of the weak equivalence sign $\approx$ appearing in the
constraints
is to remember that they are incompatible with 
the canonical Poisson brackets.}.
With this type of constraints we found more convenient 
to apply the Dirac procedure \cite{6, 7}, instead of the quicker
method proposed in Ref. \cite{fad}.

We have the four constraints
\begin{equation}
\varphi ^{\left[ 1\right] }={\cal O}^{11}\;\,,\;\,\varphi ^{\left[ 2\right]
}={\cal O}^{22}\;\,,\;\,\varphi ^{\left[ 3\right] }={\cal O}%
^{12}\;\,,\;\,\varphi ^{\left[ 4\right] }=\Lambda \;\,. 
\label{con}
\end{equation}
No secondary constraints are produced by the dynamical
compatibility condition;
thus, the complete set of constraints to deal with is given only by
Eq. (\ref{con}).
The matrix of the Poisson brackets of these constraints 
\begin{equation}
\begin{array}{l}
M^{\left[ a\right] \left[ b\right] }\left( \vec x,\vec y,t\right) =\left\{
\varphi ^{\left[ a\right] }\left( \vec x,t\right) ,\varphi ^{\left[ b\right]
}\left( \vec y,t\right) \right\} 
\medskip\  \\ \qquad \qquad \qquad \;\;=\dfrac 12\left( 
\begin{array}{cccc}
0 & 0 & \dfrac 1{2\mu } & -1 
\medskip\  \\ 0 & 0 & -\dfrac 1{2\mu } & -1 
\medskip\  \\ -\dfrac 1{2\mu } & \dfrac 1{2\mu } & 0 & 0 
\medskip\  \\ 1 & 1 & 0 & 0 
\end{array}
\right) \delta ^2\left( \vec x-\vec y\right) 
\end{array}
\end{equation}
is non-singular.

The Dirac brackets, defined by 
\begin{equation}
\begin{array}{l}
\left\{ A\left( \vec x,t\right) ,B\left( \vec y,t\right) \right\} ^{*}= 
\medskip\  \\ \qquad =\left\{ A\left( \vec x,t\right) ,B\left( \vec
y,t\right) \right\} 
\medskip\  \\ \qquad \quad \;-\dint d^2\vec z\dint d^2\vec w\left\{ A\left(
\vec x,t\right) ,\varphi ^{\left[ a\right] }\left( \vec z,t\right) \right\}
M_{\left[ a\right] \left[ b\right] }^{-1}\left( \vec z,\vec w,t\right)
\left\{ \varphi ^{\left[ b\right] }\left( \vec w,t\right) ,B\left( \vec
y,t\right) \right\} \;\,, 
\end{array}
\end{equation}
can be computed. After a
lengthy but straightforward computation, we obtain 
\begin{equation}
\left\{ k^{ij}\left( \vec x,t\right) ,k^{mn}\left( \vec y,t\right) \right\}
^{*}=\frac \mu 2\left[ {\cal E}{}^{im}\eta ^{jn}+\left( i\leftrightarrow
j\right) +\left( m\leftrightarrow n\right) \right] \delta ^2\left( \vec
x-\vec y\right) \;\,, 
\label{grabra}
\end{equation}
\begin{equation}
\left\{ h^{ij}\left( \vec x,t\right) ,k^{mn}\left( \vec y,t\right) \right\}
^{*}=\eta ^{ij}\eta ^{mn}\delta ^2\left( \vec x-\vec y\right) \;\,, 
\end{equation}
\begin{equation}
\left\{ \pi ^{0i}\left( \vec x,t\right) ,\pi ^{0j}\left( \vec y,t\right)
\right\} ^{*}=-\frac 5{64\mu }{\cal E}^{ij}\partial ^k\partial _k\delta
^2\left( \vec x-\vec y\right) \;\,, 
\end{equation}
\begin{equation}
\left\{ \pi ^{0k}\left( \vec x,t\right) ,\pi ^{mn}\left( \vec y,t\right)
\right\} ^{*}=\frac 1{64\mu ^2}{\cal E}^{ik}\left( {\cal E}^{mj}\partial ^n+%
{\cal E}^{nj}\partial ^m\right) \partial _i\partial _j\delta ^2\left( \vec
x-\vec y\right) \;\,, 
\end{equation}
\begin{equation}
\left\{ s^{ij}\left( \vec x,t\right) ,s^{mn}\left( \vec y,t\right) \right\}
^{*}=\frac 1{32\mu }\left[ {\cal E}{}^{im}\eta ^{jn}+\left( i\leftrightarrow
j\right) +\left( m\leftrightarrow n\right) \right] \delta ^2\left( \vec
x-\vec y\right) \;\,, 
\end{equation}
\begin{equation}
\left\{ \pi ^{0k}\left( \vec x,t\right) ,s^{mn}\left( \vec y,t\right)
\right\} ^{*}=\frac 1{32\mu }\left( {\cal E}{}^{km}\partial ^n+{\cal E}%
{}^{kn}\partial ^m+\eta ^{mn}{\cal E}{}^{kj}\partial _j\right) \delta
^2\left( \vec x-\vec y\right) \;\,, 
\end{equation}
\begin{equation}
\left\{ h^{\mu \nu }\left( \vec x,t\right) ,\pi ^{ij}\left( \vec y,t\right)
\right\} ^{*}=\frac 12\left( \eta ^{\mu i}\eta ^{\nu j}+\eta ^{\mu j}\eta
^{\nu i}\right) \delta ^2\left( \vec x-\vec y\right) \;\,, 
\end{equation}
\begin{equation}
\left\{ h^{ij}\left( \vec x,t\right) ,\pi ^{0k}\left( \vec y,t\right)
\right\} ^{*}=\frac 1{8\mu }\eta ^{ij}{\cal E}^{km}\partial _m\delta
^2\left( \vec x-\vec y\right) \;\,, 
\end{equation}
\begin{equation}
\left\{ h^{0\mu }\left( \vec x,t\right) ,\pi ^{0\nu }\left( \vec y,t\right)
\right\} ^{*}=\frac 12\left( \eta ^{\mu \nu }+\eta ^{\mu 0}\eta ^{\nu
0}\right) \delta ^2\left( \vec x-\vec y\right) \;\,, 
\end{equation}
\begin{equation}
\left\{ k^{ij}\left( \vec x,t\right) ,s^{mn}\left( \vec y,t\right) \right\}
^{*}=\frac 14\left( \eta ^{im}\eta ^{jn}+\eta ^{in}\eta ^{jm}-\eta ^{ij}\eta
^{mn}\right) \delta ^2\left( \vec x-\vec y\right) \;\,, 
\end{equation}
\begin{equation}
\left\{ k^{ij}\left( \vec x,t\right) ,\pi ^{0\mu }\left( \vec y,t\right)
\right\} ^{*}=\frac 18\left( \eta ^{\mu i}\partial ^j+\eta ^{\mu j}\partial
^i+3\eta ^{ij}\partial ^\mu \right) \delta ^2\left( \vec x-\vec y\right)
\;\,, 
\end{equation}
\begin{equation}
\left\{ k^{ij}\left( \vec x,t\right) ,\pi ^{mn}\left( \vec y,t\right)
\right\} ^{*}=-\frac 1{8\mu }\eta ^{ij}\left( {\cal E}^{km}\partial ^n+{\cal %
E}^{kn}\partial ^m\right) \partial _k\delta ^2\left( \vec x-\vec y\right)
\;\,.
\label{grabral}
\end{equation}
All the other brackets vanish.

The quantum theory is defined using the
generalized correspondence principle 
\begin{equation}
\left\{ A,B\right\} ^{*}\rightarrow -i\left[ A,B\right] 
\end{equation}
and defining the physical states as the ones satisfying
\begin{equation}
\left( \partial _\mu h^{\mu \nu }\right) ^{-}\left| \Phi _f\right\rangle
=0\;\,. 
\end{equation}

\section{Two-points functions}

\quad \ \thinspace 
We are particularly interested in the Wightman function, since this 
is the two point function that
enters in the definition of the scalar
product of the physical Hilbert space. In terms of 
the Wightman function $W^{(i)(j)}\left(x-y\right)$ \footnote{Henceforth, 
we shall use simplified notations.
We indicate with $(i)$ the set of all indices needed to describe
the theory; for electrodynamics, $\phi ^{(i)}=A^\mu $ and $\eta ^{(i)(j)}=\eta
^{\mu \nu }$, and for gravity, $\phi ^{(i)}=h^{\mu \nu }$ and $\eta
^{(i)(j)}=\eta ^{\mu \nu }$.
} the classical Pauli-Jordan function $\Delta
^{(i)(j)}\left( x-y\right)$, giving the canonical brackets among the
canonical variables, is expressed as
\begin{equation}
\Delta ^{(i)(j)}\left( x-y\right)
=-i\left(W^{(i)(j)}\left( x-y\right) -W^{(j)(i)}\left( y-x\right)\right) \;\,, 
\end{equation}
while the propagator is given by
\begin{equation}
S^{(i)(j)}\left( x-y\right) =\theta \left( x_0-y_0\right) W^{(i)(j)}\left(
x-y\right) +\theta \left( y_0-x_0\right) W^{(j)(i)}\left( y-x\right) \;\,. 
\label{prop}
\end{equation}

\subsection{Electrodynamics}

\quad \ \thinspace 
Consider the covariant Poisson bracket of two fields 
\begin{equation}
\Delta ^{\mu \nu }\left( x-y\right) =\left\{ A^\mu \left( x\right) ,A^\nu
\left( y\right) \right\} \;\,. 
\end{equation}
By definition, the $\Delta $ function is a solution of the second
order equations of motion 
\begin{equation}
\label{CauchyE1}\Box \Delta ^{\mu \nu }\left( z\right) +\left( \xi -1\right)
\partial ^\mu \partial _\alpha \Delta ^{\alpha \nu }\left( z\right) +\mu 
{\cal E}^{\mu \alpha \beta }\partial _\alpha \Delta _\beta {}^\nu \left(
z\right) =0\;\,. 
\end{equation}
The boundary conditions necessary for the uniqueness of
the solution stem from the canonical Poisson brackets; the equal-time
Poisson brackets correspond to the zero-time conditions for the $\Delta $
function 
\begin{equation}
\label{CauchyE2}\Delta ^{\mu \nu }\left( \vec z,0\right) =0\;\,, 
\end{equation}
\begin{equation}
\label{CauchyE3}\partial ^0\Delta ^{\mu \nu }\left( \vec z,0\right) =\left(
\eta ^{\mu \nu }+\frac{1-\xi }\xi \eta ^{\mu 0}\eta ^{\nu 0}\right) \delta
^2\left( \vec z\right) \;\,. 
\end{equation}
The $\Delta $ function is then the unique solution of the second
order Cauchy problem (\ref{CauchyE1}), (\ref
{CauchyE2}) and (\ref{CauchyE3}). Because of Lorentz covariance and 
of the symmetry properties of the Poisson
brackets, the most general form allowed for the $%
\Delta $ function is 
\begin{equation}
\Delta ^{\mu \nu }\left( z\right) =\eta ^{\mu \nu }f_1\left( z\right) +{\cal %
E}^{\mu \nu \rho }\partial _\rho f_2\left( z\right) +\partial ^\mu \partial
^\nu f_3\left( z\right) \;\,,
\label{genqed} 
\end{equation}
where $f_1$, $f_2$ and $f_3$ are scalar functions. Inserting (\ref{genqed})
into the Cauchy problem (\ref{CauchyE1}) - (\ref{CauchyE3}), 
one finds for the Wightman function  
\begin{equation}
\tilde W^{\mu \nu }\left( p\right) =2\pi \theta \left( p_0\right)
\left[ \delta \left( p^2\right) M^{(0)\alpha \beta }\left( p\right) +\delta
\left( p^2-\mu ^2\right) M^{(\mu )\alpha \beta }\left( p\right) \right]
\;\,, 
\end{equation}
with 
\begin{equation}
M^{(0)\alpha \beta }\left( p\right) =\frac i\mu {\cal E}^{\alpha \beta \rho
}p_\rho -\left( \frac 1{\mu ^2}-\frac 1\xi \frac 1{p^2}\right) p^\alpha
p^\beta \;\,, 
\end{equation}
\begin{equation}
M^{(\mu )\alpha \beta }\left( p\right) =-\left( \eta ^{\alpha \beta }-\frac{%
p^\alpha p^\beta }{\mu ^2}\right) -\frac i\mu {\cal E}^{\alpha \beta \rho
}p_\rho \;\,. 
\end{equation}
The propagator reads 
\begin{equation}
\begin{array}{l}
\tilde S^{\mu \nu }\left( p\right) =-\dfrac i{p^2+i\epsilon }\left[ -\dfrac
1\xi 
\dfrac{p^\mu p^\nu }{p^2+i\epsilon }\right] \medskip\  \\ \qquad \qquad
\;\,-\dfrac i{p^2-\mu ^2+i\epsilon }\left[ P^{\mu \nu }+i\dfrac \mu
{p^2+i\epsilon }{\cal E}^{\mu \nu \rho }p_\rho \right] \;\,, 
\end{array}
\end{equation}
with 
\begin{equation}
P^{\mu \nu }=\eta ^{\mu \nu }-\frac{p^\mu p^\nu }{p^2+i\epsilon }\;\,. 
\end{equation}

\subsection{Gravity}

\quad \ \thinspace 
For gravity one can proceed in the same way introducing the Pauli-Jordan
function in terms of the covariant Dirac bracket
\begin{equation}
\Delta^{\mu \nu \alpha \beta }\left( x-y\right) =\left\{ h^{\mu \nu }\left(
x\right) ,h^{\alpha \beta }\left( y\right) \right\} ^{*}\;\,. 
\end{equation}
By definition, the $\Delta $ function is a solution of the
equations of motion, which are in this case third order 
\begin{equation}
\label{CauchyG1} 
\begin{array}{l}
\left\{ \Box \Delta ^{\mu \nu \alpha \beta }\left( z\right) +\partial ^\mu
\partial ^\nu \Delta ^\rho {}_\rho {}^{\alpha \beta }-\eta ^{\mu \nu }\left(
\Box \Delta ^\rho {}_\rho {}^{\alpha \beta }\left( z\right) -\partial _\rho
\partial _\tau \Delta ^{\rho \tau \alpha \beta }\left( z\right) \right)
\right. 
\medskip \\ +\left( \xi -1\right) \left( \partial ^\nu \partial _\rho \Delta
^{\mu \rho \alpha \beta }\left( z\right) +\partial ^\mu \partial _\rho
\Delta ^{\rho \nu \alpha \beta }\left( z\right) \right) 
\medskip\  \\ \left. -\dfrac 1{2\mu }\left[ 
{\cal E}^{\mu \rho \tau }\partial _\rho \left( \Box \Delta _\tau {}^{\nu
\alpha \beta }\left( z\right) -\partial ^\nu \partial _\lambda \Delta
^\lambda {}_\tau {}^{\alpha \beta }\left( z\right) \right) +{\cal E}^{\nu
\rho \tau }\partial _\rho \left( \Box \Delta ^\mu {}_\tau {}^{\alpha \beta
}\left( z\right) -\partial ^\mu \partial _\lambda \Delta ^\lambda {}_\tau
{}^{\alpha \beta }\left( z\right) \right) \right] \right\} \\ \qquad =0\;\,. 
\end{array}
\end{equation}
The boundary conditions necessary to define the Cauchy problem arise
from the canonical equal-time Dirac brackets. They read
\begin{equation}
\label{CauchyG2}\Delta ^{\mu \nu \alpha \beta }\left( \vec z,0\right)
=0\;\,, 
\end{equation}
\begin{equation}
\label{CauchyG3}\partial ^0\Delta ^{ijmn}\left( \vec z,0\right) =-\eta
^{ij}\eta ^{mn}\delta ^2\left( \vec z\right) \;\,, 
\end{equation}
\begin{equation}
\label{CauchyG4}\partial ^0\Delta ^{0\mu 0\nu }\left( \vec z,0\right)
=\frac 1\xi \eta ^{\mu \nu }\delta ^2\left( \vec z\right) \;\,, 
\end{equation}
\begin{equation}
\label{CauchyG5}\partial ^0\Delta ^{ij0\mu }\left( \vec z,0\right) =0\;\,, 
\end{equation}
\begin{equation}
\label{CauchyG6}\left( \partial ^0\right) ^2\Delta ^{ijmn}\left( \vec
z,0\right) =-\frac \mu 2\left[ {\cal E}{}^{im}\eta ^{jn}+\left(
i\leftrightarrow j\right) +\left( m\leftrightarrow n\right) \right] \delta
^2\left( \vec z\right) \;\,, 
\end{equation}
\begin{equation}
\label{CauchyG7}\left( \partial ^0\right) ^2\Delta ^{ij0\mu }\left( \vec
z,0\right) =\frac 1\xi \left( \eta ^{i\mu }\partial ^j+\eta ^{j\mu
}\partial ^i+\xi \eta ^{ij}\partial ^\mu \right) \delta ^2\left( \vec
z\right) \;\,, 
\end{equation}
\begin{equation}
\label{CauchyG8}\left( \partial ^0\right) ^2\Delta ^{000\mu }\left( \vec
z,0\right) =-\frac 1\xi \partial ^\mu \delta ^2\left( \vec z\right) \;\,, 
\end{equation}
\begin{equation}
\label{CauchyG9}\left( \partial ^0\right) ^2\Delta ^{0i0j}\left( \vec
z,0\right) =0\;\,. 
\end{equation}
The $\Delta $ function is the
unique solution of the third order Cauchy problem
(\ref{CauchyG1}) - (\ref{CauchyG9}). To determine $\Delta$ explicitly
we can take again advantage of the Lorentz covariance and the symmetry
properties of the Dirac brackets to write the general form
\begin{equation}
\begin{array}{l}
\Delta ^{\mu \nu \alpha \beta }\left( z\right) =\eta ^{\mu \nu }\eta
^{\alpha \beta }f_1\left( z\right) +\left( \eta ^{\mu \alpha }\eta ^{\nu
\beta }+\eta ^{\mu \beta }\eta ^{\nu \alpha }\right) f_2\left( z\right) 
\medskip\  \\ \qquad \qquad \quad \;+\left[ \eta ^{\mu \alpha } 
{\cal E}^{\nu \beta \sigma }+\left( \mu \leftrightarrow \nu \right) +\left(
\alpha \leftrightarrow \beta \right) \right] \partial _\sigma f_3\left(
z\right) \medskip\  \\ \qquad \qquad \quad \;+\left( \eta ^{\mu \nu
}\partial ^\alpha \partial ^\beta +\eta ^{\alpha \beta }\partial ^\mu
\partial ^\nu \right) f_4\left( z\right) 
\medskip\  \\ \qquad \qquad \quad \;+\left[ \eta ^{\mu \alpha }\partial ^\nu
\partial ^\beta +\left( \mu \leftrightarrow \nu \right) +\left( \alpha
\leftrightarrow \beta \right) \right] f_5\left( z\right) 
\medskip\  \\ \qquad \qquad \quad \;+\left[ 
{\cal E}^{\mu \alpha \sigma }\partial ^\nu \partial ^\beta +\left( \mu
\leftrightarrow \nu \right) +\left( \alpha \leftrightarrow \beta \right)
\right] \partial _\sigma f_6\left( z\right) \medskip\  \\ \qquad \qquad
\quad \;+\partial ^\mu \partial ^\nu \partial ^\alpha \partial ^\beta
f_7\left( z\right) \;\,, 
\label{delta}
\end{array}
\end{equation}
where $f_1$ - $f_7$ are scalar functions. Inserting Eq. (\ref{delta}) in
the equations defining the Cauchy problem for $\Delta $,
we get, after a hard algebraic work, the Wightman function
\begin{equation}
\tilde W^{\mu \nu \alpha \beta }\left( p\right) =2\pi \theta
\left( p_0\right) \left[ \delta \left( p^2\right) M^{(0)\mu \nu \alpha \beta
}\left( p\right) +\delta \left( p^2-\mu ^2\right) M^{(\mu )\mu \nu \alpha
\beta }\left( p\right) \right] \;\,, 
\end{equation}
with
\begin{equation}
\begin{array}{l}
M^{(0)\mu \nu \alpha \beta }\left( p\right) =2\eta ^{\mu \nu }\eta ^{\alpha
\beta }-\left( \eta ^{\mu \alpha }\eta ^{\nu \beta }+\eta ^{\mu \beta }\eta
^{\nu \alpha }\right) 
\medskip \\ \qquad \qquad \qquad \;\;-\dfrac i{2\mu }\left[ \eta ^{\mu
\alpha } 
{\cal E}^{\nu \beta \sigma }+\left( \mu \leftrightarrow \nu \right) +\left(
\alpha \leftrightarrow \beta \right) \right] p_\sigma \medskip\  \\ \qquad
\qquad \qquad \;\;-\left( \dfrac 1{\mu ^2}+\dfrac 2{p^2}\right) \left( \eta
^{\mu \nu }p^\alpha p^\beta +\eta ^{\alpha \beta }p^\mu p^\nu \right) 
\medskip\  \\ \qquad \qquad \qquad \;\;+\left( \dfrac 1{\mu ^2}+ 
\dfrac{\xi -1}\xi \dfrac 1{p^2}\right) \left[ \eta ^{\mu \alpha }p^\nu
p^\beta +\left( \mu \leftrightarrow \nu \right) +\left( \alpha
\leftrightarrow \beta \right) \right] \medskip\  \\ \qquad \qquad \qquad
\;\;+\dfrac i{2\mu }\left( \dfrac 1{\mu ^2}+\dfrac 1{p^2}\right) \left[ 
{\cal E}^{\mu \alpha \sigma }p^\nu p^\beta +\left( \mu \leftrightarrow \nu
\right) +\left( \alpha \leftrightarrow \beta \right) \right] p_\sigma 
\medskip\  \\ \qquad \qquad \qquad \;\;-\left( \dfrac 1{\mu ^4}+\dfrac 1{\mu
^2p^2}-\dfrac 3\xi \dfrac 1{p^4}\right) p^\mu p^\nu p^\alpha p^\beta \;\,, 
\end{array}
\end{equation}
\begin{equation}
\begin{array}{l}
M^{(\mu )\mu \nu \alpha \beta }\left( p\right) =-\left( \eta ^{\mu \nu }- 
\dfrac{p^\mu p^\nu }{\mu ^2}\right) \left( \eta ^{\alpha \beta }-\dfrac{%
p^\alpha p^\beta }{\mu ^2}\right)\medskip \\ \qquad \qquad
\qquad \;\;\,\,+\left[ \left( \eta ^{\mu \alpha }- 
\dfrac{p^\mu p^\alpha }{\mu ^2}\right) \left( \eta ^{\nu \beta }-\dfrac{%
p^\nu p^\beta }{\mu ^2}\right) +\left( \alpha \leftrightarrow \beta \right)
\right] \medskip \\ \qquad \qquad \qquad \;\;\,\,+\dfrac i{2\mu }\left[ 
{\cal E}^{\mu \alpha \sigma }\left( \eta ^{\nu \beta }-\dfrac{p^\nu p^\beta 
}{\mu ^2}\right) +\left( \mu \leftrightarrow \nu \right) +\left( \alpha
\leftrightarrow \beta \right) \right] p_\sigma \;\,. 
\end{array}
\end{equation}

The propagator follows using Eq. (\ref{prop})
\begin{equation}
\begin{array}{l}
\tilde S^{\mu \nu \alpha \beta }\left( p\right) =-\dfrac i{p^2+i\epsilon
}\left\{ \left( P^{\mu \alpha }P^{\nu \beta }+P^{\mu \beta }P^{\nu \alpha
}\right) -2P^{\mu \nu }P^{\alpha \beta }\right. 
\medskip \\ \qquad \qquad \qquad \qquad \quad \left. +\dfrac 1\xi \left[
\eta ^{\mu \alpha } 
\dfrac{p^\nu p^\beta }{p^2+i\epsilon }+\left( \mu \leftrightarrow \nu
\right) +\left( \alpha \leftrightarrow \beta \right) \right] +\dfrac 3\xi 
\dfrac{p^\mu p^\nu p^\alpha p^\beta }{\left( p^2+i\epsilon \right) ^2}%
\right\} \medskip \\ \qquad \qquad \quad \;-\dfrac i{p^2-\mu ^2+i\epsilon
}\left[ P^{\mu \nu }P^{\alpha \beta }-\left( P^{\mu \alpha }P^{\nu \beta
}+P^{\mu \alpha }P^{\nu \beta }\right) \right. 
\medskip \\ \qquad \qquad \qquad \qquad \qquad \qquad \left. -i\dfrac \mu 2 
\dfrac{p_\sigma }{p^2+i\epsilon }\left( {\cal E}^{\mu \alpha \sigma
}P^{\beta \nu }+{\cal E}^{\nu \alpha \sigma }P^{\beta \mu }+{\cal E}^{\mu
\beta \sigma }P^{\alpha \nu }+{\cal E}^{\nu \beta \sigma }P^{\alpha \mu
}\right) \right] \;\,, 
\end{array}
\end{equation}
with, as before, 
\begin{equation}
P^{\mu \nu }=\eta ^{\mu \nu }-\frac{p^\mu p^\nu }{p^2+i\epsilon }\;\,. 
\end{equation}

\section{Quantum states}

\quad \ \thinspace 
We are now able to construct the physical Hilbert space which 
explicitly realizes the
operator algebras defined by 
Eqs.(\ref{qedbra}) and (\ref{grabra}) - (\ref{grabral}). 
This is done by defining the action of the field
operators on a suitable space of states
\footnote{The set of all regular functions defined in $O$ is 
indicated with $S\left( O\right) $. 
The $\mu $-mass-shell is called $V_\mu ^{+}$ and the corresponding
integration measure is$$
\widetilde{dp}^\mu =\left. \frac{d^2\vec p}{2\pi \sqrt{2p_0}}\right| _{p\in
V_\mu ^{+}}=\frac{d^3p}{2\pi }\sqrt{2p_0}\theta \left( p_0\right) \delta
\left( p^2-\mu ^2\right) \;\,. 
$$}.

Consider the vectorial space ${\cal H}=\left\{ \left| \Phi
\right\rangle \right\}$ of the sequences 
\begin{equation}
\left| \Phi \right\rangle =\left\{ \phi _0,\phi _1^{(\mu _1)}\left(
p_1\right) ,\phi _2^{(\mu _1)(\mu _2)}\left( p_{1,}p_2\right) ,...,\phi
_n^{(\mu _1)...(\mu _n)}\left( p_1,...,p_n\right) ,...\right\} \;\,, 
\end{equation}
where $\phi _n^{(\mu _1)...(\mu _n)}\left( p_1,...,p_n\right) \in
S\left( R^{3n}\right) $ is a completely symmetric tensor with respect to the
exchange $\left( \mu _i,p_i\right) \leftrightarrow \left( \mu _j,p_j\right)$.

The space ${\cal H}$ so defined is the Fock space, and the function $\phi
_n^{(\mu _1)...(\mu _n)}\left( p_1,...,p_n\right)$  is the $n$-particles
component of the generic state; the scalar product is defined in terms
of the Wightman function as \cite{5}
\begin{equation}
\begin{array}{l}
\left\langle \Phi \right| \left. \Psi \right\rangle
=\sum\limits_{n=0}^\infty \dint 
\dfrac{d^3p_1}{\left( 2\pi \right) ^3}...\dint \dfrac{d^3p_n}{\left( 2\pi
\right) ^3}\medskip\  \\ \qquad \qquad \quad \phi _{(\mu _1)...(\mu
_n)}^{*n}\left( p_1,..,p_n\right) \tilde W^{(\mu _1)}{}_{(\nu _1)}\left(
p_1\right) ...\tilde W^{(\mu _n)}{}_{(\nu _n)}\left( p_n\right) \psi
_n^{(\nu _1)...(\nu _n)}\left( p_1,..,p_n\right) \;\,, 
\label{scalar}
\end{array}
\end{equation}
$\forall \left| \Phi
\right\rangle $, $\left| \Psi \right\rangle \in {\cal H\ }$. 
Because of the hermiticity of the Fourier transform of the
Wightman function, the scalar product has the essential property $%
\left\langle \Phi \right| \left. \Psi \right\rangle =\left\langle \Psi
\right| \left. \Phi \right\rangle ^{*}$, $\forall \left| \Phi \right
\rangle $%
, $\left| \Psi \right\rangle \in {\cal H}$, but, as we shall see, is in
general not positive-defined.

The action of the field operators in the space of the states can be defined
using test functions.

From the positive and negative frequency parts of
the field operators $\phi ^{(i)}\left( x\right) =\phi ^{+(i)}\left( x\right)
+ \phi ^{-(i)}\left( x\right) $, one can define the
smeared operators 
\begin{equation}
\phi ^{\pm (i)}\left[ \varphi \right] =\int d^3x\phi ^{\pm (i)}\left(
x\right) \varphi \left( x\right) =\left( \phi ^{\pm (i)},\varphi \right)
\;\,.
\end{equation}
These operators have the following action on the vectors of ${\cal H}$ \cite{5}
\begin{equation}
\begin{array}{l}
\left( \phi ^{+(i)}\left[ \varphi \right] \left| \Phi \right\rangle \right)
_n^{(\mu _1)...(\mu _n)}\left( p_1,...,p_n\right) = 
\medskip\  \\ \qquad =\dfrac 1{\sqrt{n}}\sum\limits_{m=1}^n\phi
_{n-1}^{(\mu _1)...(\mu _{m-1})(\mu _{m+1})...(\mu _n)}\left(
p_1,...,p_{m-1},p_{m+1},...,p_n\right) \eta ^{(i)(\mu _m)}\tilde \varphi
\left( p_m\right) \;\,, 
\label{crea}
\end{array}
\end{equation}
\begin{equation}
\begin{array}{l}
\left( \phi ^{-(i)}\left[ \varphi \right] \left| \Phi \right\rangle \right)
_n^{(\mu _1)...(\mu _n)}\left( p_1,...,p_n\right) = 
\medskip\  \\ \qquad =\sqrt{n+1}\dint \dfrac{d^3p}{\left( 2\pi \right) ^3}%
\tilde W^{(i)}{}_{(j)}\left( p\right) \phi _{n+1}^{(j)(\mu _1)...(\mu
_n)}\left( p,p_1,...,p_n\right) \tilde \varphi \left( -p\right) \;\,. 
\label{anhi}
\end{array}
\end{equation}
By continuity, from Eqs. (\ref{crea}), (\ref{anhi}), one can obtain the
action of the unsmeared operators $\phi^{\pm(i)}\left( x\right)$;
this is usually achieved by replacing the test functions 
with Dirac $\delta$-functions, obtaining
\begin{equation}
\begin{array}{l}
\left( \phi ^{+(i)}\left( x\right) \left| \Phi \right\rangle \right)
_n^{(\mu _1)...(\mu _n)}\left( p_1,...,p_n\right) = 
\medskip\  \\ \qquad =\dfrac 1{\sqrt{n}}\sum\limits_{m=1}^n\phi
_{n-1}^{(\mu _1)...(\mu _{m-1})(\mu _{m+1})...(\mu _n)}\left(
p_1,...,p_{m-1},p_{m+1},...,p_n\right) \eta ^{(i)(\mu _m)}e^{ip_mx}\;\,, 
\end{array}
\end{equation}
\begin{equation}
\begin{array}{l}
\left( \phi ^{-(i)}\left( x\right) \left| \Phi \right\rangle \right)
_n^{(\mu _1)...(\mu _n)}\left( p_1,...,p_n\right) = 
\medskip\  \\ \qquad =\sqrt{n+1}\dint \dfrac{d^3p}{\left( 2\pi \right) ^3}%
\tilde W^{(i)}{}_{(j)}\left( p\right) \phi _{n+1}^{(j)(\mu _1)...(\mu
_n)}\left( p,p_1,...,p_n\right) e^{-ipx}\;\,. 
\end{array}
\end{equation}
With the help of these definitions, one can easily see that the field
operators satisfy the equations of motion and the covariant commutation
relations 
\begin{equation}
\left[ \phi ^{(i)}\left[ \varphi \right] ,\phi ^{(j)}\left[ \chi \right]
\right] =i\int d^3x\int d^3y\varphi \left( x\right) \Delta
^{(i)}{}^{(j)}\left( x-y\right) \chi \left( y\right) \;\,.
\label{comm} 
\end{equation}
For the unsmeared operators, Eq. (\ref{comm}) implies
\begin{equation}
\left[ \phi ^{(i)}\left( x\right) ,\phi ^{(j)}\left( y\right) \right]
=i\Delta ^{(i)}{}^{(j)}\left( x-y\right) \;\,. 
\end{equation}

Among the states contained in  ${\cal H}$ the physical states are those 
having non-vanishing
norm, $\left\langle \Phi _f\right| \left. \Phi _f\right\rangle \neq 0$, and 
satisfying the condition $G^{-}\left| \Phi _f\right\rangle =0$
\footnote{For electrodynamics,
$$
G= \partial _{\mu} A^{\mu} \;\,,
$$
while for gravity,
$$
G= \partial _{\mu} h^{\mu \nu} \;\,.
$$
}.
From the definition of the scalar product (\ref{scalar}) one can conclude
that the physical degrees of freedom of
the theory can be associated with those eigenvectors of the Wightman function
giving a non-vanishing and positive contribution to the norm and satisfying
the gauge condition.

\subsection{Electrodynamics}

\quad \ \thinspace 
Consider a one-particle state. Since the
Wightman function has domain $V_0^{+}\cup V_\mu ^{+}$, such a state
can be written as 
$$
\left| \Phi _1\right\rangle =\left\{ 0,\phi _{(0)}^\alpha \left( p\right)
+\phi _{(\mu )}^\alpha \left( p\right) ,0,0,0,...\right\} \;\,, 
$$
with 
\begin{equation}
\phi _{(0)}^\alpha \left( p\right) \in S\left( V_0^{+}\right) \ , \ \phi
_{(\mu )}^\alpha \left( p\right) \in S\left( V_\mu ^{+}\right) \;\,. 
\end{equation}
Its norm is given by 
\begin{equation}
\begin{array}{l}
\left\langle \Phi _1\right| \left. \Phi _1\right\rangle =\dint 
\widetilde{dp}^0\dfrac 1{2\pi \sqrt{2p_0}}\phi _{(0)}^{*\alpha }\left(
p\right) M_{\alpha \beta }^{(0)}\left( p\right) \phi _{(0)}^\beta \left(
p\right) \medskip \\ \qquad \qquad \quad +\dint \widetilde{dp}^\mu \dfrac
1{2\pi \sqrt{2p_0}}\phi _{(\mu )}^{*\alpha }\left( p\right) M_{\alpha \beta
}^{(\mu )}\left( p\right) \phi _{(\mu )}^\beta \left( p\right) \;\,. 
\label{norm}
\end{array}
\end{equation}

From the gauge condition $\left( \partial _\mu A^\mu \left( x\right) \right)
^{-}\left| \Phi _f\right\rangle =0$, we have 
\begin{equation}
p_\alpha \phi _{(0)}^\alpha \left( p\right) =0\;\,.
\label{gaugeqed} 
\end{equation}
From (\ref{gaugeqed}) 
\footnote{and using 
$$
\phi _{(0)}^{*\alpha }\left( p\right) {\cal E}_{\alpha \beta \rho }p^\rho
\phi _{(0)}^\beta \left( p\right) =\frac{\left( p\right) ^2}{p_0}\left(
\phi _{(0)}^{*1}\left( p\right) \phi _{(0)}^2\left( p\right) -\phi
_{(0)}^1\left( p\right) \phi _{(0)}^{*2}\left( p\right) \right) 
$$}, one can see that the massless part $\phi _{(0)}^\alpha \left(
p\right) $ of the state does not contribute to the norm (\ref{norm}), 
and, consequently, it is unphysical.

The massive part of the state appears in (\ref{norm}) only through
a projector $P_{\alpha \beta }^{(\mu )}$ according to
\begin{equation}
\left\langle \Phi _1\right| \left. \Phi _1\right\rangle =-2\int \widetilde{%
dp}^\mu \frac 1{2\pi \sqrt{2p_0}}\phi _{(\mu )}^{*\alpha }\left( p\right)
P_{\alpha \beta }^{(\mu )}\left( p\right) \phi _{(\mu )}^\beta \left(
p\right) \;\,, 
\end{equation}
with 
\begin{equation}
P^{(\mu )\alpha \beta }\left( p\right) =\frac 12\left( P^{\alpha \beta
}+\frac i\mu {\cal E}^{\alpha \beta \rho }p_\rho \right) \;\,,
\label{projqed} 
\end{equation}
\begin{equation}
P_{\alpha \rho }^{(\mu )}\left( p\right) P^{(\mu )\rho }{}_\beta \left(
p\right) =P_{\alpha \beta }^{(\mu )}\left( p\right) \;\,,\;\,P_{\alpha \beta
}^{(\mu )}\left( p\right) =P_{\beta \alpha }^{*(\mu )}\left( p\right) \;\,. 
\end{equation}
We can then conclude that the physical states have mass $\left| \mu
\right| $ and polarization given by the eigenvector of the
projector $P_{\alpha \beta }^{(\mu )}\left( p\right)$, with non vanishing
eigenvalue.

Solving explicitly the eigenvalue problem for the projector 
(\ref{projqed}), one obtains 
\begin{equation}
f^\mu \left( p\right) =\frac 1{\sqrt{2\mu ^2\left( p_0^2-\mu ^2\right) }%
}\left( 
\begin{array}{c}
\mu ^2-p_0^2 
\medskip\  \\ i\mu p^2-p^0p^1 
\medskip\  \\ -i\mu p^1-p^0p^2 
\end{array}
\right) e^{i\beta \left( p\right) }\;\,, 
\end{equation}
with the normalization 
\begin{equation}
f^{*\alpha }\left( p\right) f_\alpha \left( p\right) =-1\;\,. 
\end{equation}
The phase factor $\beta \left( p\right)$ has an 
important role and is not completely
arbitrary. As a matter of fact the infrared 
behavior of the eigenvector depends upon the
choice of the phase. Taking the infrared limit, we have 
\begin{equation}
f^\mu \left( 0\right) =\lim \limits_{\left| \vec p\right| \rightarrow 0}
\frac 1{\sqrt{2}}\left( 
\begin{array}{c}
0 
\medskip\  \\ -1 
\medskip\  \\ -i\dfrac \mu {\left| \mu \right| } 
\end{array}
\right) e^{i\left( \beta
\left( p\right) -\frac \mu {\left| \mu \right| }\theta \left( p\right)
\right) }\;\,, 
\end{equation}
with 
\begin{equation}
\theta \left( p\right) =\arctan \left( \frac{p^2}{p^1}\right) \;\,. 
\end{equation}
Since $\theta \left( p\right)$ is not defined in the origin, one must take 
\begin{equation}
\beta \left( p\right) =\frac \mu {\left| \mu \right| }\theta \left(
p\right) +\gamma \left( p\right) \;\,.
\label{bet}
\end{equation}
In (\ref{bet}) $\gamma \left( p\right) $ is an arbitrary but infrared 
regular phase factor, which can be chosen to vanish.

The final result for the polarization vector then is
\begin{equation}
\label{PolE}f^\mu \left( p\right) =\frac 1{\sqrt{2\mu ^2\left( p_0^2-\mu
^2\right) }}\left( 
\begin{array}{c}
\mu ^2-p_0^2 
\medskip\  \\ i\mu p^2-p^0p^1 
\medskip\  \\ -i\mu p^1-p^0p^2 
\end{array}
\right) e^{i\frac \mu {\left| \mu \right| }\theta \left( p\right) }\;\,. 
\label{qedeigen}
\end{equation}
The polarization vector (\ref{qedeigen}) satisfies
\begin{equation}
\label{PropE1}P^{(\mu )\mu }{}_\nu \left( p\right) f^\nu \left( p\right)
=\;\,f^\mu \left( p\right) \;\,, 
\end{equation}
\begin{equation}
\label{PropE2}p_\mu f^\mu \left( p\right) =0\;\,, 
\end{equation}
\begin{equation}
\label{PropE3}f^\mu \left( p\right) f^{*\nu }\left( p\right) =-P^{(\mu )\mu
\nu }\left( p\right) \;\,.
\end{equation}
Eqs. (\ref{PropE1}) - (\ref{PropE3}) are very useful in computing
the generators of the Poincar\'e group.
One verifies that the
contribution of the eigenvector (\ref{qedeigen}) to the norm is positive.

The phase in (\ref{qedeigen}) was first introduced in Ref.\cite{djt} as a 
regularization of the field operators. In our approach it appears directly 
in the definition of the physical states.

In conclusion, the normalized physical one-particle states of topologically 
massive electrodynamics are
\begin{equation}
\left| \Phi _{f1}\right\rangle =\left\{ 0,2\pi \sqrt{p_0}a\left( p\right)
f^\alpha \left( p\right) ,0,0,0,...\right\} \;\,.
\end{equation}
Here $a\left( p\right) $ is an arbitrary scalar function defined
on $V_\mu ^{+}$ and such that $\int d^2\vec p\left| a\left( p\right) \right|
^2=1$. $a(p)$ plays the role of the one-particle wave function.

\subsection{Gravity}

\quad \ \thinspace 
The construction of the one-particle state for the gravitational case is
very similar to that of the electrodynamics.
Since the Wightman function has
domain given by $V_0^{+}\cup V_\mu ^{+}$, this state can be written as 
$$
\left| \Phi _1\right\rangle =\left\{ 0,\phi _{(0)}^{\alpha \beta }\left(
p\right) +\phi _{(\mu )}^{\alpha \beta }\left( p\right) ,0,0,0,...\right\}
\;\,, 
$$
with 
\begin{equation}
\phi _{(0)}^{\alpha \beta }\left( p\right) \in S\left( V_0^{+}\right)\ , \
\phi _{(\mu )}^{\alpha \beta }\left( p\right) \in S\left( V_\mu
^{+}\right) \;\,. 
\end{equation}
Its norm is given by 
\begin{equation}
\begin{array}{l}
\left\langle \Phi _1\right| \left. \Phi _1\right\rangle =\dint 
\widetilde{dp}^0\dfrac 1{2\pi \sqrt{2p_0}}\phi _{(0)}^{*\alpha \beta }\left(
p\right) M_{\alpha \beta \mu \nu }^{(0)}\left( p\right) \phi _{(0)}^{\mu \nu
}\left( p\right) \medskip \\ \qquad \qquad \quad +\dint \widetilde{dp}^\mu
\dfrac 1{2\pi \sqrt{2p_0}}\phi _{(\mu )}^{*\alpha \beta }\left( p\right)
M_{\alpha \beta \mu \nu }^{(\mu )}\left( p\right) \phi _{(\mu )}^{\mu \nu
}\left( p\right) \;\,. 
\end{array}
\end{equation}

The gauge condition $\left( \partial _\mu h^{\mu \nu }\left( x\right)
\right) ^{-}\left| \Phi _f\right\rangle =0$, reads
\begin{equation}
p_\alpha \phi _{(0)}^{\alpha \beta }\left( p\right) =0\;\,.
\label{gaugegrav}
\end{equation}
Using (\ref{gaugegrav})
\footnote{and 
$$
\begin{array}{l}
\phi _{(0)}^{*\mu \nu }\left( p\right) \left[ 2\eta _{\mu \nu }\eta _{\alpha
\beta }-\left( \eta _{\mu \alpha }\eta _{\nu \beta }+\eta _{\mu \beta }\eta
_{\nu \alpha }\right) \right] \phi _{(0)}^{\alpha \beta }\left( p\right) =
\medskip\  \\ \qquad =2\left( p\right) ^2\left[ \dfrac 1{p^1p^2}\left( \phi
_{(0)}^{*01}\left( p\right) \phi _{(0)}^{02}\left( p\right) +\phi
_{(0)}^{01}\left( p\right) \phi _{(0)}^{*02}\left( p\right) \right) \right. 
\medskip\  \\ \qquad \qquad \qquad \;\;-\dfrac 1{p_0p^1}\left( \phi
_{(0)}^{*02}\left( p\right) \phi _{(0)}^{12}\left( p\right) +\phi
_{(0)}^{02}\left( p\right) \phi _{(0)}^{*12}\left( p\right) \right) 
\medskip\  \\ \qquad \qquad \qquad \;\;\left. -\dfrac 1{p_0p^2}\left( \phi
_{(0)}^{*01}\left( p\right) \phi _{(0)}^{12}\left( p\right) +\phi
_{(0)}^{01}\left( p\right) \phi _{(0)}^{*12}\left( p\right) \right) \right]
\;\,,
\end{array}
$$
$$
\begin{array}{l}
\phi _{(0)}^{*\mu \nu }\left( p\right) \left[ \eta _{\mu \alpha }
{\cal E}_{\nu \beta \sigma }+\left( \mu \leftrightarrow \nu \right) +\left(
\alpha \leftrightarrow \beta \right) \right] p^\sigma \phi _{(0)}^{\alpha
\beta }\left( p\right) 
=\medskip\  \\ \qquad 
=4\dfrac{\left( p\right) ^2}{p_0}
\left( \phi _{(0)}^{*1\lambda }\left( p\right) \phi _{(0)\lambda }^2\left(
p\right) -\phi _{(0)}^{1\lambda }\left( p\right) \phi _{(0)\lambda
}^{*2}\left( p\right) \right)
\end{array}
$$}, one sees that the massless part of the state $\phi _{(0)}^{\alpha \beta
}\left( p\right) $ does not contribute to the norm, and is
indeed unphysical.

Again the norm involves only a projection on the massive
part of the state, and reads 
\begin{equation}
\left\langle \Phi _1\right| \left. \Phi _1\right\rangle =4\int \widetilde{dp%
}^\mu \frac 1{2\pi \sqrt{2p_0}}\phi _{(\mu )}^{*\alpha \beta }\left(
p\right) P_{\alpha \beta \mu \nu }^{(\mu )}\left( p\right) \phi _{(\mu
)}^{\mu \nu }\left( p\right) \;\,, 
\end{equation}
with 
\begin{equation}
\begin{array}{l}
P^{(\mu )\mu \nu \alpha \beta }\left( p\right) =\dfrac 14\left\{ \left(
P^{\mu \alpha }P^{\nu \beta }+P^{\mu \beta }P^{\nu \alpha }\right) -P^{\mu
\nu }P^{\alpha \beta }\right. 
\medskip\  \\ \qquad \qquad \qquad \quad \;\,\left. +\dfrac i{2\mu }\left[ 
{\cal E}^{\mu \alpha \sigma }P^{\nu \beta }+\left( \mu \leftrightarrow \nu
\right) +\left( \alpha \leftrightarrow \beta \right) \right] p_\sigma
\right\} \;\,, 
\end{array}
\label{projgrav}
\end{equation}
\begin{equation}
P_{\mu \nu \rho \tau }^{(\mu )}\left( p\right) P^{(\mu )\rho \tau
}{}_{\alpha \beta }\left( p\right) =P_{\mu \nu \alpha \beta }^{(\mu )}\left(
p\right) \;\,,\;\,P_{\mu \nu \alpha \beta }^{(\mu )}\left( p\right)
=P_{\alpha \beta \mu \nu }^{*(\mu )}\left( p\right) \;\,. 
\end{equation}
We can therefore conclude that physical states have mass $\left|
\mu \right| $ and polarization given by the eigentensor of the
projector $P_{\mu \nu \alpha \beta }^{(\mu )}\left( p\right) $ with
non-vanishing eigenvalue.

Solving explicitly the eigenvalue problem for the 
projector (\ref{projgrav}), and taking
care of the infrared behavior of the eigentensor, one obtains 
\begin{equation}
\label{PolG} 
\begin{array}{l}
f^{\mu \nu }\left( p\right) =\dfrac 1{2\mu ^2}\left( 
\begin{array}{ccc}
p_0^2-\mu ^2 & p^0p^1-i\mu p^2 & p^0p^2+i\mu p^1 
\medskip\  \\ p^0p^1-i\mu p^2 & \dfrac{\left( p^0p^1-i\mu p^2\right) ^2}{%
p_0^2-\mu ^2} & \dfrac{\left( p^0p^1-i\mu p^2\right) \left( p^0p^2+i\mu
p^1\right) }{p_0^2-\mu ^2}\medskip\  \\ p^0p^2+i\mu p^1 & \dfrac{\left(
p^0p^1-i\mu p^2\right) \left( p^0p^2+i\mu p^1\right) }{p_0^2-\mu ^2} & 
\dfrac{\left( p^0p^2+i\mu p^1\right) ^2}{p_0^2-\mu ^2} 
\end{array}
\right) 
\medskip\  \\ \qquad \qquad \;e^{i2\frac \mu {\left| \mu \right| }\theta
\left( p\right) }\;\,, 
\end{array}
\end{equation}
with the normalization 
\begin{equation}
f^{*\alpha \beta }\left( p\right) f_{\alpha \beta }\left( p\right) =1\;\,. 
\end{equation}
This is the infrared well-defined polarization 
tensor that has to be used in any
perturbative computation.
This eigentensor has the following important properties 
\begin{equation}
\label{PropG1}P^{(\mu )\mu \nu }{}_{\alpha \beta }\left( p\right) f^{\alpha
\beta }\left( p\right) =f^{\mu \nu }\left( p\right) \;\,, 
\end{equation}
\begin{equation}
\label{PropG2}p_\mu f^{\mu \nu }\left( p\right) =0\;\,, 
\end{equation}
\begin{equation}
\label{PropG3}f^\mu {}_\mu \left( p\right) =0\;\,, 
\end{equation}
\begin{equation}
\label{PropG4}f^{\mu \nu }\left( p\right) f^{*\alpha \beta }\left( p\right)
=P^{(\mu )\mu \nu \alpha \beta }\left( p\right) \;\,. 
\end{equation}
One can verify that the contribution of the eigentensor (\ref{PolG}) to the
norm is positive.

Summarizing, the normalized physical one-particle states of topologically
massive gravity are  
\begin{equation}
\left| \Phi _{f1}\right\rangle =\left\{ 0,\pi \sqrt{2p_0}a\left( p\right)
f^{\alpha \beta }\left( p\right) ,0,0,0,...\right\} \;\,, 
\end{equation}
where $a\left( p\right) $ is an arbitrary scalar function defined
on $V_\mu ^{+}$ and such as $\int d^2\vec p\left| a\left( p\right) \right|
^2=1$. This plays the role of one-particle wave function.

Note that the eigentensor of gravity
is the tensorial product of two eigenvectors of electrodynamics: 
\begin{equation}
f^{\mu \nu }\left( p\right) =f^\mu \left( p\right) f^\nu \left( p\right)
\;\,. 
\end{equation}
Mathematically, this stems from the fact that on $V_\mu ^{+}$ one
has 
\begin{equation}
P^{(\mu )\mu \nu \alpha \beta }\left( p\right) =\frac 12\left( P^{(\mu )\mu
\alpha }\left( p\right) P^{(\mu )\nu \beta }\left( p\right) +P^{(\mu )\mu
\beta }\left( p\right) P^{(\mu )\nu \alpha }\left( p\right) \right) \;\,. 
\end{equation}
Physically, the two theories essentially differ in the spin
representation to which they correspond. Since the spin
is a scalar quantity in 2+1 dimensions, and since the two
theories correspond to representations of the Poincar\'e group which are
related by a tensorial product (as expected \cite{djt}), 
the spin of the graviton will be twice that
of the photon. We shall show explicitly in the next section that the states
constructed above satisfy this property. 

\section{Observables}

\quad \ \thinspace 
The last step in the analysis of the two theories is to
extract the physical properties of the states. This is readily done
by expanding the fields in normal modes and defining creation and
annihilation operators for the physical states.

We start by observing that a field satisfying the equations of motion can be
expanded in normal modes in a completely general way, once the
Wightman function of the theory is known. In fact, one can write 
\begin{equation}
\phi ^{(i)}\left( x\right) =\int \frac{d^3p}{\left( 2\pi \right) ^2}\sqrt{%
2p_0}\left[ \tilde W^{(i)}{}^{(j)}\left( p\right) a_{(j)}\left( p\right)
e^{-ipx}+\tilde W^{*(i)}{}^{(j)}\left( p\right) a_{(j)}^{+}\left( p\right)
e^{ipx}\right] \;\,. 
\end{equation}
According to the spectral theorem, the Wightman function can
always be written as
\begin{equation}
\tilde W^{(i)}{}^{(j)}\left( p\right) =2\pi \theta \left( p_0\right)
\sum_{n=1}^N\delta \left( p^2-\mu _n^2\right) M^{\mu _n(i)}{}^{(j)}\left(
p\right) \;\,, 
\end{equation}
where the $\mu _n$ have the dimension of a mass, the $M^{\mu
_n(i)}{}^{(j)}\left( p\right) $ are hermitian matrices and $N$ is the number
of massive and massless excitation of the theory
\footnote{As already seen, in both our cases we have $\mu _1=0$ and 
$\mu _2=\mu $.}. The field expansion can then be written as 
\begin{equation}
\phi ^{(i)}\left( x\right) =\sum\limits_{n=1}^N\int \widetilde{dp}^{\mu
_n}\left[ M^{\mu _n(i)}{}^{(j)}\left( p\right) a_{(j)}^{\mu _n}\left(
p\right) e^{-ipx}+M^{*\mu _n(i)}{}^{(j)}a_{(j)}^{+\mu _n}\left( p\right)
e^{ipx}\right] \;\,. 
\end{equation}
Using the eigenvectors of the $N$ matrices $M^{\mu
_n(i)}{}^{(j)}\left( p\right) $, defined by the eigenvalues problem 
\begin{equation}
M^{\mu _n(i)}{}_{(j)}\left( p\right) f_{\left[ k\right] }^{\mu _n(j)}\left(
p\right) =\lambda _{\left[ k\right] }^{\mu _n}\left( p\right) f_{\left[
k\right] }^{\mu _n(i)}\left( p\right) \;\,,\;\,p\in V_{\mu _n}^{+}\;\,, 
\end{equation}
and normalized to 
\footnote{Since the internal field metric $\eta^{(i)(j)}$ is not in 
general positive defined, one has to consider both signs.}
\begin{equation}
f_{\left[ k\right] }^{\mu _n(i)}\left( p\right) f_{\left[ l\right]
(i)}^{*\mu _n}\left( p\right) =\pm \eta _{\left[ k\right] \left[ l\right]
}\;\,,\;p\in V_{\mu _n}^{+}\;\,, 
\end{equation}
the expansion finally reads 
\begin{equation}
\phi ^{(i)}\left( x\right) =\sum\limits_{n=1}^N\sum\nolimits_{\left[
k\right] }\int \widetilde{dp}^{\mu _n}\sqrt{\left| \lambda _{\left[ k\right]
}^{\mu _n}\left( p\right) \right| }\left[ a_{\left[ k\right] }^{\mu
_n}\left( p\right) f_{\left[ k\right] }^{\mu _n(i)}\left( p\right)
e^{-ipx}+a_{\left[ k\right] }^{+\mu _n}\left( p\right) f_{\left[ k\right]
}^{*\mu _n(i)}\left( p\right) e^{ipx}\right] \;\,. 
\end{equation}

The explicit action of the operators $a_{\left[ k\right] }^{\mu _n}\left[
\varphi \right] $ and $a_{\left[ k\right] }^{+\mu _n}\left[ \varphi \right] $
on the states can be deduced from that of the fields; the smeared
operators 
\begin{equation}
a_{\left[ k\right] }^{\mu _n}\left[ \varphi \right] =\int d^2\vec pa_{\left[
k\right] }^{\mu _n}\left( p\right) \varphi \left( p\right) =\left( a_{\left[
k\right] }^{\mu _n},\varphi \right) \;\,, 
\end{equation}
\begin{equation}
a_{\left[ k\right] }^{+\mu _n}\left[ \varphi \right] =\int d^2\vec
pa_{\left[ k\right] }^{+\mu _n}\left( p\right) \varphi ^{*}\left( p\right)
=\left( a_{\left[ k\right] }^{+\mu _n},\varphi ^{*}\right) \;\,, 
\end{equation}
act like 
\begin{equation}
\begin{array}{l}
\left( a_{\left[ k\right] }^{+\mu _p}\left[ \varphi \right] \left| \Phi
\right\rangle \right) _n^{(\mu _1)...(\mu _n)}\left( p_1,...,p_n\right) = 
\medskip\  \\ \qquad =\pm 
\dfrac{2\pi \sqrt{2p_0}}{\sqrt{n}}\sum\limits_{m=1}^n\phi _{n-1}^{(\mu
_1)...(\mu _{m-1})(\mu _{m+1})...(\mu _n)}\left(
p_1,...,p_{m-1},p_{m+1},...,p_n\right) \medskip\  \\ \qquad \quad \dfrac{%
\eta _{\left[ k\right] \left[ k\right] }}{\sqrt{\left| \lambda _{\left[
k\right] }^{\mu _p}\left( p_m\right) \right| }}f_{\left[ k\right] }^{\mu
_p(\mu _m)}\left( p_m\right) \varphi ^{*}\left( p_m\right) \;\,, 
\end{array}
\end{equation}
\begin{equation}
\begin{array}{l}
\left( a_{\left[ k\right] }^{\mu _p}\left[ \varphi \right] \left| \Phi
\right\rangle \right) _n^{(\mu _1)...(\mu _n)}\left(p_1,...,p_n\right) = 
\medskip\  \\ \qquad =\pm \sqrt{n+1}\dint \widetilde{dp}^{\mu _p}\eta
_{\left[ k\right] \left[ k\right] }\dfrac{\lambda _{\left[ k\right] }^{\mu
_p}\left( p\right) }{\sqrt{\left| \lambda _{\left[ k\right] }^{\mu _p}\left(
p\right) \right| }}f_{\left[ k\right] (i)}^{*\mu _p}\left( p\right) \phi
_{n+1}^{(i)(\mu _1)...(\mu _n)}\left( p,p_1,...,p_n\right) \varphi \left(
p\right) \;\,. 
\end{array}
\end{equation}
It then follows that the unsmeared operators act like
\begin{equation}
\begin{array}{l}
\left( a_{\left[ k\right] }^{+\mu _p}\left( p\right) \left| \Phi
\right\rangle \right) _n^{(\mu _1)...(\mu _n)}\left( p_1,...,p_n\right) = 
\medskip\  \\ \qquad =\pm 
\dfrac{2\pi \sqrt{2p_0}}{\sqrt{n}}\sum\limits_{m=1}^n\phi _{n-1}^{(\mu
_1)...(\mu _{m-1})(\mu _{m+1})...(\mu _n)}\left(
p_1,...,p_{m-1},p_{m+1},...,p_n\right) \medskip\  \\ \qquad \quad \;\dfrac{%
\eta _{\left[ k\right] \left[ k\right] }}{\sqrt{\left| \lambda _{\left[
k\right] }^{\mu _p}\left( p_m\right) \right| }}f_{\left[ k\right] }^{\mu
_p(\mu _m)}\left( p_m\right) \delta ^2\left( \vec p-\vec p_m\right) \;\,, 
\end{array}
\end{equation}
\begin{equation}
\begin{array}{l}
\left( a_{\left[ k\right] }^{\mu _p}\left( p\right) \left| \Phi
\right\rangle \right) _n^{(\mu _1)...(\mu _n)}\left( p_1,...,p_n\right) = 
\medskip\  \\ \qquad =\pm \dfrac{\sqrt{n+1}}{2\pi \sqrt{2p_0}}\eta _{\left[
k\right] \left[ k\right] }\dfrac{\lambda _{\left[ k\right] }^{\mu _p}\left(
p\right) }{\sqrt{\left| \lambda _{\left[ k\right] }^{\mu _p}\left( p\right)
\right| }}f_{\left[ k\right] (i)}^{*\mu _p}\left( p\right) \phi
_{n+1}^{(i)(\mu _1)...(\mu _n)}\left( p,p_1,...,p_n\right) \;\,. 
\end{array}
\end{equation}
The commutation relations of $a_{\left[ k\right] }^{\mu _n}\left[
\varphi \right] $ and $a_{\left[ k\right] }^{+\mu _n}\left[ \varphi \right] $
 are 
\begin{equation}
\left[ a_{\left[ k\right] }^{\mu _m}\left[ \varphi \right] ,a_{\left[
l\right] }^{+\mu _n}\left[ \chi \right] \right] =\pm \delta ^{mn}\eta
_{\left[ k\right] \left[ l\right] }\frac{\left| \lambda _{\left[ k\right]
}^{\mu _m}\left( p\right) \right| }{\lambda _{\left[ k\right] }^{\mu
_m}\left( p\right) }\int d^2\vec p\varphi \left( p\right) \chi ^{*}\left(
p\right) \;\,, 
\end{equation}
\begin{equation}
\left[ a_{\left[ k\right] }^{\mu _m}\left[ \varphi \right] ,a_{\left[
l\right] }^{\mu _n}\left[ \chi \right] \right] =\left[ a_{\left[ k\right]
}^{+\mu _m}\left[ \varphi \right] ,a_{\left[ l\right] }^{+\mu _n}\left[ \chi
\right] \right] =0\;\,, 
\end{equation}
which imply 
\begin{equation}
\left[ a_{\left[ k\right] }^{\mu _m}\left( p\right) ,a_{\left[ l\right]
}^{+\mu _n}\left( q\right) \right] =\pm \delta ^{mn}\eta _{\left[ k\right]
\left[ l\right] }\frac{\left| \lambda _{\left[ k\right] }^{\mu _m}\left(
p\right) \right| }{\lambda _{\left[ k\right] }^{\mu _m}\left( p\right) }%
\delta ^2\left( \vec p-\vec q\right) \;\,, 
\label{alg1}
\end{equation}
\begin{equation}
\left[ a_{\left[ k\right] }^{\mu _m}\left( p\right) ,a_{\left[ l\right]
}^{\mu _n}\left( q\right) \right] =\left[ a_{\left[ k\right] }^{+\mu
_m}\left( p\right) ,a_{\left[ l\right] }^{+\mu _n}\left( q\right) \right]
=0\;\,.
\label{alg2}
\end{equation}
Thus, $a_{\left[ k\right] }^{+\mu _n}\left( p\right) $ and $a_{\left[
k\right] }^{\mu _n}\left( p\right) $ are creation and annihilation operators
for the states of mass $\left| \mu _n\right| $ and polarization $f_{\left[
k\right] }^{\mu _n(i)}\left( p\right) $. The physical Fock space can be
constructed by the cyclic action of $a_{\left[ k\right]
}^{+\mu _n}\left( p\right) $ on the vacuum.

In the next subsections we shall compute, on the physical states,
the mean-values of the Poincar\'e charges 
\begin{equation}
P^\mu =\int d^2\vec xT^{0\mu }\;\,,\; \ M^{\mu \nu }=\int d^2\vec
xM^{0\mu \nu }\;\,, 
\end{equation}
and of the Pauli-Lubanski scalar, which defines the spin through 
\begin{equation}
S=\frac 1{2\left| \mu \right| }{\cal E}_{\alpha \mu \nu }P^\alpha M^{\mu \nu
}\;\,. 
\end{equation}
Then, by means of the algebra (\ref{alg1}), (\ref{alg2}), we shall 
verify that the Poincar\'e
is satisfied on the space of the physical states.

\subsection{Electrodynamics}

\quad \ \thinspace 
The physical field is 
\begin{equation}
\label{FieldE}A_f^\mu \left( x\right) =\sqrt{2}\int \widetilde{dp}^\mu
\left[ a\left( p\right) f^\mu \left( p\right) e^{-ipx}+a^{+}\left( p\right)
f^{*\mu }\left( p\right) e^{ipx}\right] \;\,. 
\end{equation}
From the properties of $f^\mu \left( p\right) $, one obtains for the physical
field
\begin{equation}
\label{ConsE1}A_f^\mu \left( x\right) -\frac 1\mu {\cal E}^{\mu \alpha
\beta }\partial _\alpha A_{f\beta }\left( x\right) =0\;\,, 
\end{equation}
\begin{equation}
\label{ConsE2}\partial _\mu A_f^\mu \left( x\right) =0\;\,, 
\end{equation}
so that the physical excitation has Klein-Gordon dynamics 
\begin{equation}
\left( \Box +\mu ^2\right) A_f^\mu \left( x\right) =0\;\,. 
\end{equation}

The complete Poincar\'e generators contain terms depending also on the
unphysical components (the massless and the massive
with vanishing norm) of the field. Our analysis enables to
select only the physical part of the Poincar\'e generators, which are given by 
\begin{equation}
P_f^\mu =\int d^2\vec pp^\mu a^{+}\left( p\right) a\left( p\right) \;\,, 
\end{equation}
\begin{equation}
M_f^{ij}={\cal E}^{ij}\dint d^2\vec p\left[ a^{+}\left( p\right) \left( 
\overleftrightarrow{-\frac i2\frac \partial {\partial \theta }}\right)
a\left( p\right) +\frac \mu {\left| \mu \right| }a^{+}\left( p\right)
a\left( p\right) \right] \;\,, 
\end{equation}
\begin{equation}
M_f^{0i}=tP_f^i+\dint d^2\vec p\left[ a^{+}\left( p\right) \left( 
\overleftrightarrow{\dfrac i2p^0\partial ^i}\right) a\left( p\right) +\dfrac
\mu {\left| \mu \right| }\dfrac 1{\left| \mu \right| +p^0}{\cal E}%
^i{}_kp^ka^{+}\left( p\right) a\left( p\right) \right] \;\,, 
\end{equation}
and satisfy the Poincar\'e algebra.

On a physical state 
\begin{equation}
\left| \Phi _{f1}\left( k\right) \right\rangle =a^{+}\left( k\right) \left|
\Phi _{f0}\right\rangle \;\,, 
\end{equation}
one has 
\begin{equation}
\left\langle \Phi _{f1}\left( k\right) \right| P_f^\mu \left| \Phi
_{f1}\left( k\right) \right\rangle =k^\mu \;\,, 
\end{equation}
\begin{equation}
\left\langle \Phi _{f1}\left( k\right) \right| S_f\left| \Phi _{f1}\left(
k\right) \right\rangle =\frac \mu {\left| \mu \right| }\;\,. 
\end{equation}
Therefore, $a^{+}\left( k\right) $ creates a photon with mass $%
\left| \mu \right| $, four-momentum $k^\mu $, spin $\frac \mu {\left| \mu
\right| }$ and polarization $f^\mu \left( k\right)$. This is the physical
excitation of the theory already found in Ref. \cite{djt}.

It is important to stress that the algebra realization exhibited in this
paper, is the only one compatible with the closure of
the Poincar\'e algebra. The phase choice performed in Section 4,
not only makes the polarization vector infrared
well-behaved, but it is also the phase choice that allows the Poincar\'e
algebra to close. If we did not include the regularization phase in the
definition of the $f^\mu(p)$, the Poincar\'e algebra would present the 
well-known anomaly obtained in \cite{djt}. In fact the generators become 
\begin{equation}
P_f^\mu =\int d^2\vec pp^\mu a^{+}\left( p\right) a\left( p\right) \;\,, 
\end{equation}
\begin{equation}
M_f^{ij}={\cal E}^{ij}\dint d^2\vec p a^{+}\left( p\right) \left( 
\overleftrightarrow{-\frac i2\frac \partial {\partial \theta }}\right)
a\left( p\right) \;\,, 
\end{equation}
\begin{equation}
M_f^{0i}=tP_f^i+\dint d^2\vec p\left[ a^{+}\left( p\right) \left( 
\overleftrightarrow{\dfrac i2p^0\partial ^i}\right) a\left( p\right) +
\frac{{\cal E}^i{}_kp^k}{p^mp_m} a^{+}\left( p\right) a\left( p\right) 
\right] \;\,, 
\end{equation}
and the Poincar\'e algebra acquires
the anomaly 
\begin{equation}
\Delta=2\pi\mu^2 S a^{+}(0)a(0)
\end{equation}
in the commutator of two boosts
\begin{equation}
\left[ M_f^{0i},M_f^{0j}\right] =-i\left( M_f^{ij}-{\cal E}^{ij}\Delta
\right) \;\,.
\end{equation}
Consequently, in any phenomenological computation, if we
want a true representation of the Poincar\'e algebra, we need to use the
polarization vector (\ref{PolE}). 

\subsection{Gravity}

\quad \ \thinspace 
The physical field is 
\begin{equation}
\label{FieldG}h_f^{\mu \nu }\left( x\right) =2\int \widetilde{dp}^\mu \left[
a\left( p\right) f^{\mu \nu }\left( p\right) e^{-ipx}+a^{+}\left( p\right)
f^{*\mu \nu }\left( p\right) e^{ipx}\right] \;\,. 
\end{equation}
From the properties of $f^{\mu \nu }\left( p\right)$
follow the properties of the physical field
\begin{equation}
\label{ConsG1}h_f^{\mu\nu }\left(p\right)-\frac 1{2\mu }\left[ {\cal E}%
^{\mu \alpha \beta }\partial _\alpha h_{f\beta }{}^\nu \left( x\right) +%
{\cal E}^{\nu \alpha \beta }\partial _\alpha h_{f\beta }{}^\mu \left(
x\right) \right] =0\;\,, 
\end{equation}
\begin{equation}
\label{ConsG2}\partial _\mu h_f^{\mu \nu }\left( x\right) =0\;\,, 
\end{equation}
\begin{equation}
\label{ConsG3}h_f^\mu {}_\mu \left( x\right) =0\;\,, 
\end{equation}
 and the physical excitation has Klein-Gordon dynamics 
\begin{equation}
\left( \Box +\mu ^2\right) h_f^{\mu \nu }\left( x\right) =0\;\,. 
\end{equation}

The physical Poincar\'e generators are those of the
electrodynamic case, except for a factor two in the spin terms. The Poincar\'e
algebra is satisfied and $a^{+}\left( k\right)  $ creates a graviton with mass 
$\left| \mu\right| $, four-momentum $k^\mu $, 
spin $2\frac \mu {\left| \mu \right| }$
and polarization $f^{\mu \nu }\left( k\right) $ \cite{djt}. As before, the
spin dependent phase choice performed in Section 4, to regularize the
infrared behavior of the polarization tensor, allows the closure of the
Poincar\'e algebra. 

\section{Conclusion}

\quad \ \thinspace 
The detailed analysis of the free quantum states of the topologically massive
theories performed in this paper, has provided us with the explicit form of the
one-particle states for these theories. We have shown that the states are
infrared well-behaved representations of the Poincar\'e group with the well
known values for the physical charges (mass and spin). 
These states define the polarization 
vector and tensor needed in phenomenological
computations. The phase, regularizing the infrared
behavior of the physical states, 
allows also for the closure of the Poincar\'e
algebra on the physical states.

The derivation of the electrodynamical physical states might have 
relevant consequences on the knowledge of the bound-state spectrum of
the theory.
In particular the existence of photon-fermion bound-states suggested in
Ref. \cite{3}, might be proved by the analysis of the effective 
non-relativistic 
fermion-photon potential. In fact, the inclusion of the regularizing phase
in the photon polarization vector, simplifies dramatically the angular
dependence of such a potential.
We shall discuss how the regularization phase
affects the bound-state spectrum of
the theory in a forthcoming paper \cite{gss}. 

For gravity it would be interesting to compare our results
with an analogous
treatment of the gauge theoretical first order 
formulation\cite{4}. The comparison of the
free quantum states of the two theories  might provide
important informations on the relation between the first order and metric
gravity at the quantum level. 

\section{Acknowledgments}

\quad \ \thinspace 
We are grateful to M. Mintchev and G. Nardelli for useful discussions.

\end{document}